\documentstyle[ltwol,epsf]{article}

\input FEYNMAN

\def\be{\begin{equation}}
\def\ee{\end{equation}}
\def\bea{\begin{eqnarray}}
\def\\eea{\end{eqnarray}}

\begin{document}

\title{Leptoproduction of $J/\psi$}

\author{Thomas Mehen}

\address{California Institute of Technology, Pasadena, CA 91125
\\E-mail: mehen@theory.caltech.edu}   

\twocolumn[\maketitle\abstracts{I review the status of the extraction of NRQCD
color-octet $J/\psi$ production matrix elements. Large theoretical uncertainties
in current extractions from hadropoduction and photoproduction are emphasized. 
Leptoproduction of $J/\psi$ is calculated within the NRQCD factorization
formalism. Color-octet contributions dominate the cross section, allowing for a
reliable extraction of $\langle {\cal O}_8^{J/\psi}(^1S_0)\rangle$ and $\langle
{\cal O}_8^{J/\psi}(^3P_0)\rangle$. A comparison with preliminary data from the
H1 collaboration shows that leading order color-octet mechanisms agree with the
measured total cross section for $Q^2 > 4m_c^2$, while the color-singlet model
underpredicts the cross section by a factor of 5. }]

The NRQCD factorization formalism of Bodwin, Braaten, and Lepage\cite{BBL} has
emerged as a new paradigm for computing the production and decay rates of heavy
quarkonia. This formalism provides a rigorous theoretical framework which
systematically incorporates relativistic corrections and ensures the infrared
safety of perturbative calculations.\cite{IR} In the NRQCD factorization
formalism, cross sections for the production of a quarkonium state $H$ are
written as
\begin{equation}
\label{Factor}
\sigma(H) =
\sum_n {c_n(\alpha_s, m_Q) \over m^{d_{n}-4}_{Q}} 
\langle 0|{\cal O}_n^H|0\rangle ,
\end{equation}
where $m_{Q}$ is the mass of the heavy quark $Q$.
The short-distance coefficients, $c_n$, are associated with the
production, at distances of order $1/m_{Q}$ or less, of a $Q\bar{Q}$ pair 
with quantum numbers indexed by $n$ (angular momentum, ${}^{2S+1}L_{J}$,
and color, $1$ or $8$). They are computable in perturbation theory.
In Eq.~(\ref{Factor}), $\langle 0| {\cal O}_n^H|0 \rangle$ are  
vacuum matrix elements of NRQCD operators:
\begin{equation}
\label{On}
\langle 0| {\cal O}_n^H|0 \rangle \equiv \sum_X \sum_{\lambda}
\langle 0|{\cal K}_n^{\dagger}|H(\lambda) + X\rangle
\langle H(\lambda) + X |{\cal K}_n|0 \rangle,
\end{equation}
where ${\cal K}_n$ is a bilinear in heavy quark fields which creates a $Q
\bar{Q}$ pair in an angular-momentum and color configuration indexed by $n$. 
The bilinear combination ${\cal K}_n^{\dagger}{\cal K}_n$ has energy dimension
$d_{n}$.  The production matrix elements describe the evolution of the $Q
\bar{Q}$ pair into a final state containing the quarkonium $H$ plus additional
hadrons ($X$) which are soft in the quarkonium rest frame. Throughout the
remainder of this talk,  a shorthand notation will be used in which the vacuum
matrix elements are written as $\langle {\cal O}^H_{(1,8)}(^{2S+1}L_J) \rangle$.

The NRQCD matrix elements obey simple scaling laws\cite{lmnmh} with respect to
$v$, the relative velocity of the $Q$ and $\bar{Q}$ in the quarkonium bound
state. Therefore, Eq.~(\ref{Factor}) is a double expansion in $v$ and
$\alpha_s$. Since the NRQCD matrix elements are sensitive only to large distance
scales, they are independent of the short-distance process in which the $Q$ and
$\bar{Q}$ are produced. Thus, the NRQCD matrix elements are universal parameters
which can be extracted from one experiment and used to predict production cross
sections in other processes.

Prior to the innovations presented in Ref.~\cite{BBL}, most $J/\psi$ production
calculations took into account only the hadronization of $c\bar{c}$ pairs
initially produced in a color-singlet ${}^3S_1$ state, as parameterized by the
NRQCD matrix element $\langle{\cal O}^{J/\psi}_1({}^3S_1)\rangle$.  An important
aspect of the NRQCD formalism is that, in addition to the color-singlet
contribution, it allows for the possibility that a $c\overline{c}$ pair produced
in a color-octet state can evolve nonperturbatively into a $J/\psi$. The most
important color-octet matrix elements are $\langle{\cal O}^{J/\psi}_8({}^3S_1)
\rangle$, $\langle{\cal O}^{J/\psi}_8({}^1S_0) \rangle$, and $\langle{\cal
O}^{J/\psi}_8({}^3P_J) \rangle$, which are suppressed by $v^4$ relative to the
leading color-singlet matrix element.  They describe the non-perturbative
evolution of a color-octet $c\overline{c}$ pair in either a ${}^3S_1$,
${}^1S_0$, or ${}^3P_J$ angular momentum state into a $J/\psi$.  Using heavy
quark spin symmetry relations, it is possible to express all three
P-wave matrix elements in terms of one: $\langle {\cal O}_8(^3P_J)\rangle =
(2J+1) \langle {\cal O}_8(^3P_0)\rangle + O(v^2)$. Thus, at this order in the
$v$ expansion, there are three independent color-octet matrix elements.

The NRQCD factorization formalism has enjoyed considerable phenomenological
success.  Most notably, the large $p_\perp$ production of $J/\psi$ at hadron
colliders, which is underpredicted by over an order of magnitude in the
color-singlet model (CSM), can be easily accounted for by color-octet production
mechanisms.\cite{bf} The CSM cannot account for the observed branching ratios
for $B \to J/\psi + X$, $B \to \psi^{\prime} +X$. Inclusion of color-octet
mechanisms removes this discrepancy.\cite{kls,bmr} Color-octet mechanisms also
improve the understanding of $J/\psi$ production in $Z^0$ decay.\cite{cho} 
  
The NRQCD factorization formalism has yet to be conclusively proven as the
correct theory of quarkonium production. One important test of the formalism is
the verification of the universality of the NRQCD matrix elements. Another
important prediction, which has yet to be verified, is the polarization of
$J/\psi$ at large $p_\perp$ at hadron colliders. Large $p_\perp$ production of
$J/\psi$ is dominated by fragmentation of a (nearly) on-shell gluon into a
$^3S_1^{(8)}$ $c{\bar c}$ pair, which inherits the gluon's transverse
polarization. Because of heavy quark spin symmetry, soft gluons emitted as the
$c{\bar c}$ pair hadronizes into the quarkonium state do not dilute this
polarization.\cite{cw,br}  Therefore, at large $p_\perp$, $J/\psi$ are expected
to be almost 100\% transversely polarized. This gets significantly diluted at
lower $p_\perp$ due to nonfragmentation production.\cite{al,bk} 

Polarization is a particularly interesting test of quarkonium production
because it can distinguish between NRQCD and the other remaining model of
quarkonium production, the color evaporation model (CEM).\cite{wisc} In this
model, the cross section for producing a $J/\psi$ is proportional to the total
production rate for $c\bar{c}$ pairs with invariant mass less than the
$D\bar{D}$ threshold:
\begin{eqnarray}
\sigma_{J/\psi} = \rho_{J/\psi} \int_{2 m_c}^{2 m_D} dm_{c\bar{c}} {d \sigma_{c\bar{c}} \over d m_{c\bar{c}} },
\end{eqnarray}
where $\rho_{J/\psi}$ is a universal factor which represents the fraction of
$c\bar{c}$ pairs which hadronize into a $J/\psi$. Because this model includes
color-octet production mechanisms, it is consistent with hadron collider data as
well as data from $Z^0$ decay\cite{wisc}. A prediction of this model is that the
emission of soft gluons in the hadronization washes out any polarization of the
$c\bar{c}$ produced in the short distance process. This prediction is at odds
with heavy quark spin symmetry arguments. The CEM would be a reasonable model if
the relative velocity, $v$, in the quarkonium state were too large to serve as a
useful expansion parameter. From the existing estimates of the color-octet
matrix elements, this does not appear to be the case. The leading color-singlet
matrix element is well determined: $\langle {\cal O}^{J/\psi}_1(^3S_1) \rangle =
1.1 \pm 0.1/, {\rm GeV}^3$. Extractions of $\langle {\cal O}^{J/\psi}_8(^3S_1)
\rangle, \langle {\cal O}^{J/\psi}_8(^1S_0) \rangle$, and $\langle {\cal
O}^{J/\psi}_8(^3P_0) \rangle/m_c^2$ suffer from considerable uncertainty but
they are known to be of order $10^{-2}\,{\rm GeV}^3$. These values are
consistent with the $v^4$ suppression expected on the basis of NRQCD scaling
laws.  

To quantitatively predict polarization as a function of $p_\perp$, color-octet
matrix elements need to be accurately determined. At the present time,
extractions from production at hadron colliders suffer from considerable
theoretical uncertainty. Beneke and Kr\"amer\cite{bk} extract $\langle {\cal
O}^{J/\psi}_8(^3S_1) \rangle = 1.2^{+1.2}_{-0.7} \times 10^{-2}\, {\rm GeV}^3
$, where statistical errors, and errors due to variation of the renormalization
scale and the parton distribution functions, have been added in quadrature. The
error is dominated by scale uncertainty. Other extractions\cite{cl,cgmp} of
$\langle {\cal O}^{J/\psi}_8(^3S_1) \rangle$ are consistent within these
errors.  $\langle {\cal O}^{J/\psi}_8(^1S_0) \rangle, \langle {\cal
O}^{J/\psi}_8(^3P_0) \rangle$ are even more poorly determined since
hadroproduction is only sensitive to a particular linear combination. The
extracted value of this linear combination is also much more sensitive to the
choice of parton distribution function and the effects of intital and final
state radiation.\cite{kk,ccsl} Extractions from lower energy
hadroproduction\cite{br2} suffer similar from problems.  These calculations are
extremely sensitive to the charm quark mass because of the small energy scale
of the process.  Corrections due to higher orders in perturbation theory and
higher twist effects are expected to be large.

Recently, progress has been made in analysis of $B$ decays and $Z^0$ decay. 
A next-to-leading order analysis of $B$ decay\cite{bmr} establishes a bound on
the following linear combination of color-octet matrix elements:
\begin{eqnarray}
\langle {\cal O}^{J/\psi}_8(^1S_0)\rangle + 3.1 {\langle {\cal O}^{J/\psi}_8(^3P_0)\rangle \over m_c^2} < 2.3 \times 10^{-2} \,{\rm GeV}^3 .
\end{eqnarray}
The individual matrix elements, $\langle O^{J/\psi}_8(^1S_0)\rangle$ and
$\langle {\cal O}^{J/\psi}_8(^3P_0)\rangle$ remain undetermined. Production of
$J/\psi$ in $Z^0$ decays is dominated by gluon fragmentation and is therefore 
sensitive to $\langle {\cal O}^{J/\psi}_8(^3S_1)\rangle$. A recent
analysis,\cite{lr} which sums logarithms of $2 M_{J/\psi}/M_Z$ and $2
E_{J/\psi}/M_Z$ to all orders, extracts the following effective matrix element:
\begin{eqnarray}
\langle {\hat{\cal O}}^{J/\psi}_8(^3S_1)\rangle & = & \sum_H \langle {\hat{\cal O}}^{H}_8(^3S_1)\rangle {\rm Br}(H \to J/\psi+X) \nonumber \\
& = & 1.9 \pm 0.5 \pm 1.0\times 10^{-2} \,{\rm GeV}^3 .
\end{eqnarray}
The first error is statistical, the second theoretical. The effective $\langle
{\hat{\cal O}}^{J/\psi}_8(^3S_1)\rangle$ includes feeddown from higher mass
charmonium resonances ($H$) because it is not possible to seperate prompt
$J/\psi$ experimentally.

Next-to-leading order calculations of photoproduction would seem to indicate
that NRQCD predicts a large excess of $J/\psi$ at large $z$ which is not
observed in the data.\cite{ck} (Here $z= E_{J/\psi}/E_{\gamma}$, where energies
are measured in the proton rest frame.) This led to the conclusion that
the color-octet matrix elements extracted from the Tevatron were an order of
magnitude too large to be consistent with photoproduction data from the HERA
experiments.  However, this enhancement is actually an artifact of the
next-to-leading order perturbative calculation.  It is worth examining this
process in more detail because it is indicative of some of the pitfalls one may
encounter when trying to compare NRQCD with data.

\begin{figure}
\epsfxsize=8cm
\hfil\epsfbox{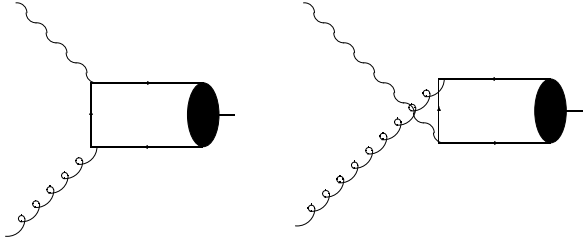}\hfill
\caption{Leading order diagrams for color-octet photoproduction of
$J/\psi$}
\label{born}
\end{figure}

The leading order diagrams are shown in Fig.\ \ref{born}, and results in the 
following perturbative cross section:
\begin{eqnarray}
\label{lo}
{d \sigma \over d z} &=& \sigma_0 \delta (1-z), \\
\sigma_0 &=&  {4 \pi^3 \alpha_s \alpha \over s_{\gamma p} m_c^3}G(4 m_c^2/s_{\gamma p}) \times\\
&& \left(\langle {\cal O}^{J/\psi}_8(^1S_0)\rangle + 7 {\langle
 {\cal O}^{J/\psi}_8(^3P_0)\rangle \over m_c^2} \right) \nonumber 
\end{eqnarray}
$G(x_g)$ is the gluon distribution function of the proton. $s_{\gamma p}$ is
the center of mass energy squared of the photon-proton system.  The delta
function in Eq.(\ref{lo}) is an artifact of the perturbative expansion which
will be smeared by nonperturbative emission of soft gluons as the $c\bar{c}$
pair hadronizes into the final state hadron. How this nonperturbative smearing
can be accounted for within the NRQCD formalism will be discussed below. For
now, we concern ourselves with perturbative corrections.

\begin{figure}
\epsfxsize=8cm
\hfil\epsfbox{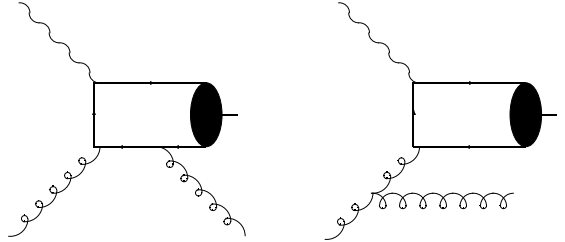}\hfill
\caption{Next-to-leading order diagrams for color-octet photoproduction of
$J/\psi$}
\label{nlodiagrams}
\end{figure}

The next-to-leading order diagrams are shown in Fig.\ \ref{nlodiagrams}. 
In the limit in which the final state gluon is soft, the cross section is given by: \cite{mmp} 
\begin{eqnarray}
\label{nlo}
{d \sigma \over d z} &= &\int dx_g G(x_g) {\sum |M_{LO}|^2 \over 16 \pi x_g s_{\gamma p}} \nonumber \\
&& \times C_A g_s^2 \left( {2P\cdot g \over P\cdot k g \cdot k} - {4m_c^2 \over (P\cdot k)^2} \right) \nonumber \\
&=&\int_\rho^z dx G(\rho/x) {\sum |M_{LO}|^2 \over 4 m_c^2 s_{\gamma p}} C_A \alpha_s {z-x \over (1-x)^2 (1-z)}  \nonumber \\
&=&-\sigma_0 {C_A \alpha_s \over \pi}{{\rm ln}(1-z) \over 1-z} + ...
\end{eqnarray}
$M_{LO}$ is the leading order matrix element from Fig.\ \ref{born}. $P$, $g$,
and $k$ are the four momentum of the $J/\psi$, the initial state gluon, and the
final state gluon, respectiverly. $\rho = 4 m_c^2/s_{\gamma p}$ and
$x=\rho/x_g$. In the last line of Eq.(\ref{nlo}), we take $x=1$ inside the
argument of the structure function in order to obtain an analytic expression.
This approximation is valid up to subleading logs since the leading logs come
from the region $x \to z \to 1$.  In higher orders, we expect to find terms of
the form $\alpha_s^n {\rm ln}^{2n-1}(1-z)/(1-z)$ which need to be resummed
before theory can be compared with experimental data. The divergence of the
cross section as $z \to 1$ is an artifact of the next-to-leading order
calculation. Once the resummation is performed, the cross section will be
well-behaved as $z \to 1$.

There are also important nonperturbative corrections\cite{Shape} near $z=1$. 
For quarkonium production near the boundaries of phase space, it is sometimes
the case that contributions from NRQCD operators which are higher order in $v$
are enhanced by kinematic factors.  This results in the breakdown of the NRQCD
expansion.  The crux of the problem is that in the perturbative QCD part of the
matching calculation one uses twice the heavy quark mass instead of the
quarkonium mass to compute the phase space for the production of the quarkonium
meson. The difference between $2m_{Q}$ and $M_{H}$ is a $v^{2}$ correction,
which is ignored in leading-order calculations. However, at the boundaries of
phase space this difference becomes important, and it is necessary to sum an
infinite number of NRQCD matrix elements.  This resummation leads to a universal
distribution function called a shape function which replaces the delta function
in Eq.\ (\ref{lo}).  Because of the universality of the shape functions, it may
be possible in the future to test NRQCD by comparing shape functions extracted
from different quarkonium production processes. This would require more precise
data than is currently available.  

As $z \to 1$, the photoproduction cross section becomes sensitive to the effects
of soft gluon radiation. Within pertubation theory this sensitivity is signalled
by the presence of large logarithms which must be resummed. The $v$ expansion in
Eq.\ (\ref{Factor}) also breaks down and an infinite set of operators become
relevant. For this reason, it is impossible to extract NRQCD matrix elements
from this distribution. When leptoproduction is discussed below, the analysis
will focus on observables for which shape function corrections are not
kinematically enhanced, because it is only from these distributions where one
can hope to obtain a reliable extraction of NRQCD matrix elements.

In Ref.\cite{fm}, leptoproduction of $J/\psi$ is examined and found to be a
useful process from which to extract NRQCD matrix elements. This process is
similar to photoproduction, however, now the photon is off-shell. At leading
order, the cross section is again given by the diagrams in Fig.\ \ref{born}: 
\begin{eqnarray}
\lefteqn{ {d \sigma \over d Q^2} = \int dy \int dx_g\; G(x_g)\;
\delta(x_g ys-(2m_c)^{2}-Q^2) } \nonumber \\
 & & \;\;\;\;\; \times {2 \alpha_s(\mu) \alpha^2 e^2_c \pi^2 \over Q^2 
(Q^{2}+(2m_{c})^{2})m_c} \Bigg\{{1+(1-y)^2 \over y} \Big[
 \langle{\cal O}^{\psi}_8(^1S_0)\rangle \nonumber \\
 & & \;\;\;\;\;\;\;\;\;\;\;\;\;\;+ {3Q^2+7(2m_c)^2 \over Q^{2}+(2m_{c})^{2}}
 {\langle{\cal O}^{\psi}_8(^3P_0)\rangle \over m^2_c} \Big] \nonumber \\
 & & \;\;\;\;\;\;\;\;\;\;\;\;\;\; -{8(2m_c)^2Q^2 \over x_g s (Q^{2}+(2m_{c})^{2})}{\langle{\cal O}^{\psi}_8(^3P_0)\rangle \over m^2_c} \Bigg\} \; ,
 \label{epcs}
\end{eqnarray} 
where $s$ is the electron-proton center-of-mass energy squared.  The momentum
fraction of the virtual photon relative to the incoming lepton is $y \equiv
P_{p}\cdot q / P_{p} \cdot k$, where $P_{p}$ is the proton four-momentum, $q$ is
the photon four-momentum, and $k$ is the incoming lepton four-momentum, and
$Q^{2} \equiv -q^{2}$. 

There are several advantages to extracting NRQCD matrix elements from this
process. First of all, note that the relative importance of $\langle {\cal
O}^{\psi}_8(^1S_0)\rangle$ and $\langle {\cal O}^{\psi}_8(^3P_0)\rangle$ changes
as a function of $Q^2$.  Thus, it is possible to fit the differential cross
section as a function of $Q^2$ and extract both of these matrix elements
seperately. Second, the scale for the coupling is set by $\mu^2 \approx Q^2
+(2m_c)^2$, so as $Q^2$ increases, perturbative corrections should be
increasingly suppressed relative to photoproduction. In Ref.\cite{fm},
next-to-leading order graphs with real gluon emission were computed and found to
give a small contribution to the total cross section for $Q^2 > 4\, {\rm
GeV}^2$. The uncertainty due to varying the renormalization scale is less than
$10\%$ for $Q^2 > 4\, {\rm GeV}^2$. This is a great deal smaller than the
corresponding uncertainty in hadroproduction.  This indicates that higher
order perturbative corrections to the leading order color-octet production
mechanism will be small.  Higher twist corrections to the parton model are also
expected to be small as they are suppressed by powers of $Q^2$.

The leading color-singlet mechanism requires the emission of an additional hard
gluon. (See the first diagram in Fig.\ \ref{nlodiagrams}.) This $\alpha_s(Q)$
suppression compensates for the $v^4$ suppression of the color-octet matrix
elements, and it turns out that for $Q^2 > 4\, {\rm GeV}^2$, the color-octet
contribution is roughly 4 times bigger than the leading color-singlet
process. However, color-singlet mechanisms will dominate in the inelastic ($z
<1$) region, where there is no $O(\alpha_s)$ color-octet contribution. In the
elastic region, one also expects production of $J/\psi$ via diffractive
processes. At large $Q^2$, diffractive leptoproduction can be studied using
perturbative QCD. In Ref. \cite{Diffractive}, a perturbative analysis of
diffractive leptoproduction predicts the cross section to fall as $1/(Q^2 + (2
m_c)^2)^3$, as compared to $1/(Q^2 + (2 m_c)^2)^2$ for the color-octet
mechanism. Therefore, at sufficiently large $Q^2$, the diffractive contribution
should be negligible correction to our calculation.

One must also consider the possibility of kinematic enhancement of higher order
$v^2$ corrections. An analysis of the $Q^2$ distribution in Ref.\ \cite{fm}
shows that the higher order in $v^2$ corrections associated with the shape
function are suppressed in the large  $Q^2$ limit.  Therefore, this distribution
is calculable in NRQCD. However, the $z$ distribution still suffers from large
endpoint corrections, and a perturbative calculation of this distribution cannot
be compared with experiment near $z = 1$, even at large $Q^2$.  

The largest errors in the calculation of the $Q^2$ distribution are associated
with uncertainty in the charm quark mass. Varying $m_c$ between $1.3 \,{\rm
GeV}$ and $1.7 \,{\rm GeV}$, results in an error of $^{+60\%}_{-25 \%}$ at $Q^2 =
10~{\rm GeV}^2$. This error decreases slightly as $Q^2$ is increased.

Once color-octet matrix elements are extracted, the polarization of the $J/\psi$
can be predicted without introducing any new parameters. Calculations of the
polarization of $J/\psi$ produced via the leading order color-octet mechanism as
a function of $Q^2$ are given in Ref.\ \cite{fm}.  A precise measurement of the
polarization of leptoproduced $J/\psi$ could provide an excellent test of the
NRQCD factorization formalism.  

Preliminary results from HERA,\cite{H1} indicate that $d\sigma/d Q^2$ is well
described by NRQCD for $Q^2 > (2 m_c)^2$. This is in contrast with CSM
calculations which underpredict the cross section by  a factor of $\sim 5$. So
this appears to be strong evidence in favor of color-octet mechanisms as
understood within the NRQCD factorization formalism. However, the shape of the
rapidity distribution at large rapidity is not well reproduced by NRQCD.  This
could be a consequence of the effects of soft gluon emission which are not
accounted for in the leading order calculations of Ref.\ \cite{fm}.  Analysis of
soft gluon effects and extraction of NRQCD matrix elements will be performed in
a future publication.\cite{fm2}

\section*{Acknowledgements}

This work was done in collaboration with Sean Fleming. This work was supported
by the National Science Foundation under grant number PHY-94004057, by the
Departmnent of Energy under grant number DE-FG03-92-ER40701, and by a John A.
McCone Fellowship.

\end{document}